\documentclass[letterpaper, 12pt]{ewshm}
\pdfoutput=1
\usepackage{scalerel}
\usepackage{graphicx}
\usepackage{color}
\usepackage{amsfonts}
\usepackage{amsmath}
\usepackage{amssymb}
\usepackage{times}
\usepackage{cite}
\usepackage{hhline}
\usepackage{compactbib}
\usepackage{subcaption}
\usepackage{booktabs, tabularx}
\newcolumntype{C}{>{\centering\arraybackslash}X}
\usepackage{draftwatermark}
\SetWatermarkText{Draft}
\SetWatermarkScale{1.5}
\usepackage[utf8]{inputenc}
\AtEndDocument{\par\leavevmode}

\renewcommand{\thefigure}{\arabic{figure}}
\renewcommand{\thetable}{\Roman{table}}

\long\def\symbolfootnote[#1]#2{\begingroup%
\def\thefootnote{\fnsymbol{footnote}}\footnote[#1]{#2}\endgroup}

\makeatletter
\renewcommand\@biblabel[1]{#1.}
\makeatother

\begin{document}

 \vspace{\baselineskip}
 \vspace{\baselineskip}
\begin{center}
	\Large{\textbf{On the Potential of Dynamic Substructuring Methods for Model Updating}}\\
	\normalsize
	\vspace{\baselineskip}
	\vspace{\baselineskip}
	\vspace{\baselineskip}
	\vspace{\baselineskip}
	Thomas Simpson$^{1}$, Vasilis Dertimanis$^{1}$, Costas Papadimitriou$^{2}$and Eleni Chatzi$^{1}$\\ 
	\vspace{\baselineskip}
	$^{1}$ Institute of Structural Engineering, Department of  Civil, Environmental and Geomatic Engineering, ETH Zürich, Stefano-Franscini Platz 5, 8093, Zürich, Switzerland\\
	$^{2}$ C Papadimitriou, Department of Mechanical Engineering, University of Thessaly, Volos 38334, Greece\\
\end{center}
\normalsize
\vspace{\baselineskip}
\vspace{\baselineskip}
\vspace{\baselineskip}

\section*{ABSTRACT}
\label{abstract}
While purely data-driven assessment is feasible for the first levels of the Structural Health Monitoring (SHM) process, namely damage detection and arguably damage localization, this does not hold true for more advanced processes. The tasks of damage quantification and eventually residual life prognosis are invariably linked to availability of a representation of the system, which bears physical connotation. In this context, it is often desirable to assimilate data and models, into what is often termed a digital twin of the monitored system.

One common take to such an end lies in exploitation of structural mechanics models, relying on use of Finite Element approximations. proper updating of these models, and their
incorporation in an inverse problem setting may allow for damage quantification and localization, as well as more advanced tasks, including reliability analysis and fatigue assessment. However, this may only be achieved by means of repetitive analyses of the forward model, which implies considerable computational toll, when the model used is a detailed FE representation. In tackling this issue, reduced order models can be adopted, which retain the parameterisation and link to the parameters regulating the physical properties, albeit greatly reducing the computational burden.

In this work a detailed FE model of a wind turbine tower is considered, reduced forms of this model are found using both the Craig Bampton and Dual Craig Bampton methods. These reduced order models are then used and compared in a Transitional Markov Chain Monte Carlo procedure to localise and quantify damage which is introduced to the system.

%------------------------------------------------------------------------------------------------------------
% SECTION 1: INTRODUCTION
%------------------------------------------------------------------------------------------------------------

\vspace{12pt} 
\newpage
\noindent \uppercase{\textbf{Introduction}}  \vspace{12pt} 

Within the hierarchy of structural health monitoring, problems of damage detection and to some extent damage localisation have been shown to be amenable to analysis using purely data based pattern recognition algorithms \cite{Peeters2001},\cite{Dervilis2014}. However, when considering higher levels in the hierarchy such as damage assessment and even performance prognosis, these purely statistical methods are currently found to be lacking without extensive physical testing \cite{MANSON2003}. As such, physics based models are required when considering these higher order problems.

A common technique is to use parameterised finite element models which can predict certain dynamic properties of the system of interest. The parameters of the model can be updated to match the dynamic properties of interest using various model-updating techniques \cite{Jafarkhani2011},\cite{Teughels2002}. The variation in these parameters, which often represent quantities such as local stiffness of a part of the model, can localise and effectively quantify damage as a loss of local stiffness.

For many structures of interest however, these finite element models can regularly consist of hundreds of thousands of degrees of freedom. This poses a significant problem for many model-updating techniques, which can often require a large number of re-analyses of the FE model especially for the popular Bayesian MCMC method. This issue has inspired the previous use of various model reduction techniques which retain the parameterisation and prediction of required dynamic properties of the original model whilst greatly reducing the computational burden of re-analyses \cite{Jensen2014},\cite{Papadimitriou2013}.

Previous studies have separately made use of dynamic substructuring techniques inspired by structural dynamics \cite{Papadimitriou2013},\cite{Jensen2014}. Within this study, a comparison is made between traditional dynamic substructuring techniques including the Craig-Bampton (CB) \cite{Craig1968} and Dual Craig-Bampton (DCB) \cite{Rixen2004} methods.

In this work, a wind turbine tower is considered which is parametrised by local Young’s modulus values. An initial undamaged state model is used to create the parameterised reduction bases for the CB and DCB methods. Damage is then introduced into the system by locally reducing the elastic modulus to simulate a decrease in stiffness as occurs in the presence of a crack. Following this, a model updating procedures is followed using a Transitional Markov Chain Monte Carlo procedure \cite{Ching}. This is carried out for each of the two reduced order model variants. Each of these reduced order models is compared according to the accuracy of the updated parameters and the computational burden of the update process.

The model updating procedure is shown to effectively both locate and quantify the damage present in the system, whilst the incurred computational burden is greatly reduced via exploitation of the model reduction schemes, which are here comparatively assessed.

\vspace{24pt} 
\noindent \uppercase{\textbf{Case Study Description}} \vspace{12pt} 

The model considered herein is a section of wind turbine tower constructed of a linear elastic steel material. The complete model consists of $34839$ DOFs and is meshed using tetrahedral elements of $6$ DOF. The tower section has a total height of 40 m, with an outer diameter tapering from 8 m at the base to 5 m at the top, the wall thickness of the tower section is 0.1 m and the flange which is located at the midpoint has an outer diameter of 6.6 m and a total thickness of 0.5 m. The material properties of the tower are assumed to be a Youngs Modulus of 210 GPa and a Poissons ratio of 0.3.

The model is considered to be made up of two components, an upper and lower section, which are joined at a flange. This joint is assumed to act as a rigid connection. Damage in the structure is modelled as a local decrease in stiffness, whilst each of the substructures is characterised by a single uniform stiffness parameter.
\begin{figure}[h!]
    \centering
    \begin{subfigure}[b]{0.3\textwidth}
    \centering
        \includegraphics[height=60mm]{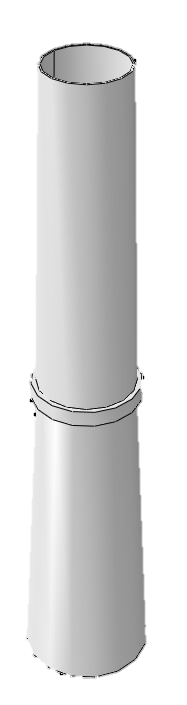}
        \caption{Assembled Tower}
        \label{fig:gull}
    \end{subfigure}
    \quad
    \begin{subfigure}[b]{0.3\textwidth}
    \centering
        \includegraphics[height=60mm]{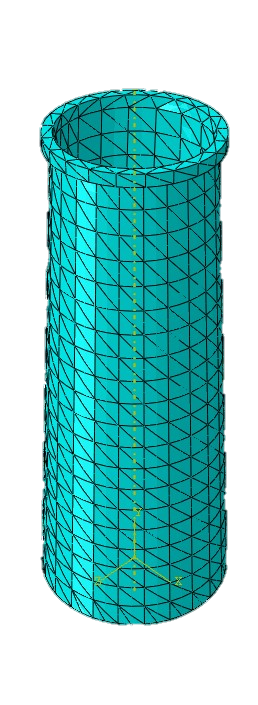}
        \caption{Lower Section}
        \label{fig:tiger}
    \end{subfigure}
    \quad
    \begin{subfigure}[b]{0.3\textwidth}
    \centering
        \includegraphics[height=60mm]{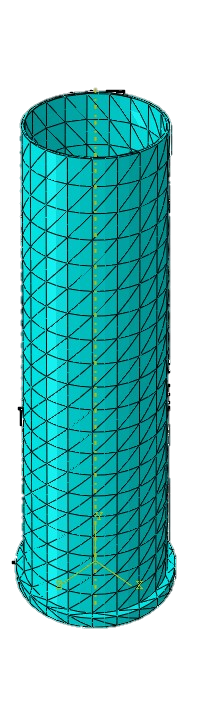}
        \caption{Upper Section}
        \label{fig:mouse}
    \end{subfigure}
    \caption{Finite element models of the wind turbine tower considered}\label{fig:animals}
\end{figure}

\vspace{12pt} 
\noindent \uppercase{\textbf{Finite Element Model Updating}} \vspace{12pt} 

The possibility to effectively update finite element models using data is fundamental to their real-life applicability. The use of numerical models in forward mode, i.e., without a calibration to real life structures may only offer very limited and unreliable insight. Model updating is typically performed by i) characterising the model using certain parameters, which are to be updated, ii) comparing the results of the computer model to results produced from a real life structure, and iii) eventually updating these parameters so that the results closely approximate the monitored response \cite{Jaishi2005},\cite{Friswell1995}. This parameter updating step can be carried out in several ways, with a popular framework being the use of Bayesian MCMC methods.

The use of model updating for damage detection comprises a slight extension of this concept. This typically is a two-step process, with the first step requiring definition of an approximately calibrated numerical model for a monitored structure. In a second step, the discrepancy between the estimates of the numerical model and the real life structure are utilised for picking up changes in the state of the structure and hence as indications of the possible presence of damage. As an added step, an additional model updating procedure can be carried out, using parameters which may now represent the effect of damage onto the structure. A common choice in this respect are stiffness parameters given that damage, typically expressed in the form of crack growth, can have associated effects of a decrease in stiffness. Furthermore, if these parameters are local then beyond damage detection, the tasks of localisation and quantification may further be achieved \cite{Teughels2002}.

In order to carry out model updating, a function must be specified which can quantify the discrepancy between the measured results and the numerical model. This would be the residual function in a typical optimisation framework, or the likelihood in a Bayesian context. Within this study, the values of interest are the eigenvalues and eigenvectors of the measured system and the numerical model, corresponding to the commonly adopted procedure of modal identification. More precisely, the objective function J is expressed relative to the error between the measured and calculated eigenvalues, and the modal assurance criterion (MAC) between the measured and calculated eigenvectors \cite{Pastor2012}.

\begin{equation}
    f_{err}=\frac{1}{N}\sum_{i=1}^{n}{\frac{(f_i-f_{y,i})^2}{f_{y,i}^2}},\quad
        MAC_{err}=\frac{1}{NM}\begin{bmatrix}
    
   \frac{|\Phi_t^T\Phi_r|^2}{(\Phi_t^T\Phi_t)(\Phi_r^T\Phi_r)}-I
    \end{bmatrix}^2
\end{equation}
%\begin{equation}
 %   MAC_{err}=\frac{1}{NM}\begin{bmatrix}
    
  % \frac{|\Phi_t^T\Phi_r|^2}{(\Phi_t^T\Phi_t)(\Phi_r^T\Phi_r)}-I
  %  \end{bmatrix}^2
%\end{equation}
\begin{equation}
    J=f_{err}+MAC_{err}
\end{equation}

The Bayesian formulation of the model updating problem involves finding the parameter values, which maximise the posterior distribution of the parameter values given the model structure, the measured data, and the prior assumption on the parameter values.
\begin{equation}
    P(\theta|D,M)=\frac{P(D|\theta,M)P(\theta|M)}{P(D|M)}
\end{equation}
In this case, the prior values are assinged a log normal distribution with their mean value set at 1. The log-normal distribution was chosen as it is a strictly greater than zero distribution, which is logical for stiffness values. The mean of the distribution was set at 1 as this indicates the value of no change in stiffness, i.e., the initial assumption that no damage is present.

There are various options for algorithms to perform the Bayesian learning operation. In this case the Transitional Markov Chain Monte Carlo (TMCMC) was used. This is a variation on more standard MCMC algorithms, which has been shown to handle sampling from complex posterior distributions better than standard methods \cite{Ching}.

\vspace{24pt}
\noindent \uppercase{\textbf{Dynamic Substructuring}} \vspace{12pt} 

Dynamic substructuring (DS) is principally concerned with the decomposing of large dynamic models into several smaller simpler models and the subsequent implementation of a coupling regime between the substructures to represent the larger model. The key motivation behind DS techniques is that it allows for seperate analysis of each of the substructures; this can greatly reduce the computational burden of a given dynamic analysis problem. This seperation can also allow for specific analysis methods for different regions. For example, some substructures can be analysed linearly whilst others are considered non-linearly or even allowing for a combination of numerical and experimental substructures \cite{Voormeeren2011}.
%Furthermore, it can allow for specialised analysis techniques for each of the substructures in turn including combination of experimentally derived and computational models \cite{Voormeeren2011}.

DS is often used in combination with model reduction techniques in the form of component mode synthesis (CMS) methods. The traditional variants rely upon model reduction by using a modal decomposition of the substructures, from which only a subset of contributing modes is maintained. This method is popular in structural dynamics due to the lower frequency modes being heavily dominant in global response \cite{Craig1985}.

The application of DS methods herein concerns the reduction of the computational effort of model re-analyses, such that the Bayesian identification of parameters becomes feasible without use of large or high performance computing assets. To this end, it is required that the model can be separated into sub-structures each of which is characterised by a parameter, such as stiffness, whose variation is linked to possible damage. Each sub-structure is then individually reduced wherein it is important that the reduced form can be parameterised according to these local stiffness parameters. The DS methods exploited in this work include the Craig-Bampton method and the Dual Craig-Bampton method.

\vspace{12pt} 
 
\noindent \textbf{Craig-Bampton}  \vspace{12pt} 

The Craig-Bampton is one of the most prominent and long standing techniques in CMS and is already implemented in numerous commercial finite element softwares \cite{Craig1968}. 
In the Craig-Bampton method each sub-structure can be reduced individually and then the reduced sub-structures are coupled together. The key aspects of the reduction are that the degrees of freedom (DOFs) of each substructure are partitioned into into internal DOFs and external DOFs, i.e., those which are on the interface between substructures.
\begin{equation}
    \begin{bmatrix}
    M_{ii} & 
    M_{ib} \\
    M_{bi} & 
    M_{bb}
    \end{bmatrix}
    \ddot{
    \begin{bmatrix}x_{i}\\x_b
    \end{bmatrix}
    }+
    \begin{bmatrix}
    K_{ii}&K_{ib}\\
    K_{bi}&K_{bb}
    \end{bmatrix}
    \begin{bmatrix}
    x_{i}\\x_b
    \end{bmatrix}=0
\end{equation}
The internal DOFs, which in most cases are the majority, can be significantly reduced using a modal decomposition from the eigenvalue problem in Equation \ref{eq:inMod}. Given that for structural dynamics the lowest eigenmodes are dominant, the majority of these modes can be discarded, with only the lower modes retained.
 
In addition to the fixed interface normal modes, constraint modes are also used to characterise the boundary DOFs and the static response. Constraint modes correspond to the static deformation shape due to a unit displacement applied at a boundary DOF. As such there exist as many, as there exist boundary DOFs. These are calculated as in Equation \ref{eq:constraintmodes}.
\begin{equation}\label{eq:constraintmodes}
    \Psi=-K_{ii}^{-1}K_{ib}
\end{equation}

Using both the fixed interface and constraint modes, the original high dimensional coordinate set can be approximated on a significantly lower coordinate set, with the reduction of DOFs dependant upon the number of fixed interface modes that are discarded.

\begin{equation}
\begin{bmatrix}
x_i\\x_b
\end{bmatrix}
\approx
\begin{bmatrix}
    \Phi_r & \Psi
    \\0& I
\end{bmatrix}
\begin{bmatrix}
q\\x_b
\end{bmatrix}
=R
\begin{bmatrix}
q\\x_b
\end{bmatrix}
\end{equation}
The mass and stiffness matrices of the sub-structures can then be reduced by projection onto these reduced co-ordinate sets.
\begin{equation}
    \hat{K}=R^TKR
    \quad   
    \hat{M}=R^TMR
\end{equation}

The above process can be carried out to individually reduce each of the separate sub-structures. As a last step, the global response should be reconstructed from the individual substructures. In the case of the CB method a primal assembly approach is used. This entails enforcing compatibility by equating the corresponding boundary DOFs of each substructure, hence it naturally requires conforming meshes. First, the reduced matrices are assembled in block diagonal matrices. 
\begin{equation}
    \tilde{K}=\begin{bmatrix}
    \hat{K_1}&0
    \\0&\hat{K_2}
    \end{bmatrix}
    \quad
    \tilde{M}=\begin{bmatrix}
    \hat{M_1}&0
    \\0&\hat{M_2}
    \end{bmatrix}
\end{equation}
The primal assembly matrix can then be formed as shown in Equation \ref{eq:primalass}. This matrix is used to enforce the compatability constraint between the boundary DOFs.
\begin{equation}\label{eq:primalass}
    L=\begin{bmatrix}
    I_{i1}&0&0\\
    0&0&I_{b}\\
    0&I_{i2}&0\\
    0&0&I_{b}
    \end{bmatrix}
\end{equation}
The final reduced and coupled matrices representing the global system can then be found. 
\begin{equation}
    K_{CB}=L^T\tilde{K}L \quad M_{CB}=L^T\tilde{M}L
\end{equation}
The final eigenproblem is hence reduced to a significantly lower dimensionality problem. The number of eigenpairs which can be extracted is naturally also greatly reduced, however the low frequency modes are still retained to an accuracy level dependent upon the extent of the reduction.
\begin{equation}
    (K_{CB}-\lambda M_{CB})=0
\end{equation}

\noindent \textbf{Dual Craig-Bampton}  \vspace{12pt} 

The Dual Craig Bampton (DCB) method is a relatively recently developed CMS method which whilst currently much less widely used, comes with numerous advantages \cite{Rixen2004},\cite{inproceedings}. Several studies demonstrate better fidelity of the DCB method than the conventional CB method for a given number of retained modes.

The DCB method uses a dual assembly to couple the substructures as opposed to the primary assembly method used in the traditional CB. Dual assembly involves coupling substructures by enforcing equal and opposite coupling forces between boundary DOFs in the substructures. This results in both substructures retaining all of their DOFs in the coupling procedure. As such, it is possible to use reduction basis for the substructures which rely on free interface normal modes extracted from the full set of DOFs.

The interface assembly forces are included in the equations of motion for each sub-structure as shown in Equation \ref{eq:dualass}. These interface forces are constructed by way of a vector of Lagrange multipliers $\mu$ and matrix $B$, which is a boolean matrix, which defines the equal and opposite forces between boundary DOFs. The compatability condition is in the Dual case explicitly enforced by equation \ref{eq:comp} as it is not apriori fulfilled as in primal assembly.
\begin{equation}\label{eq:dualass}
    M\ddot{u}+Ku+B\mu=f
\end{equation}
\begin{equation}\label{eq:comp}
    Bu=0
\end{equation}

This assembly can be applied to the mass and stiffness matrices in block diagonal form and the equations are assembled as in Equation \ref{eq:Dualeqs}.

\begin{equation}
\hat{M}=
\begin{bmatrix}
M_1&0\\0&M_2
\end{bmatrix}
\quad
\hat{K}=
\begin{bmatrix}
K_1&0\\0&K_2
\end{bmatrix}
\end{equation}

\begin{equation}\label{eq:Dualeqs}
    \begin{bmatrix}
    \hat{M}&0\\0&0
    \end{bmatrix}
    \begin{bmatrix}
    \ddot{u}\\ \mu
    \end{bmatrix}
    +
    \begin{bmatrix}
    \hat{K}&B\\B^T&0
    \end{bmatrix}
    \begin{bmatrix}
    u\\ \mu 
    \end{bmatrix}
    =
    \begin{bmatrix}
    f\\0
    \end{bmatrix}
\end{equation}
Given this assembly regime each substructure can then be reduced individually and in all DOFs. The dynamic behaviour of the systems is reduced in terms of an eigendecomposition of the full matrices from which low frequency modes and any present rigid body modes can be retained.
\begin{equation}
M_i^{-1}K_i=\Phi_i\Sigma_i\Phi_i^T
\end{equation}
\begin{equation}
    \Phi_i=
    \begin{bmatrix}
    R_i&\Phi_{i,r}&\Phi_{i,d}
    \end{bmatrix}
\end{equation}

In addition to the free interface normal modes, attachment modes are also used. These modes define the response of the substructure to the interface forces used for coupling. They correspond to the deformation of the substructure under a unit force at one of the boundary DOF and hence are related to the flexibility matrix.

More precisely, given the possible presence of rigid body modes in the substructures, a generalised inverse of the stiffness matrix is taken. In the presence of rigid body modes these are projected out of the flexibility matrix. Given that the flexibility matrix can be formed as a modal decomposition, the contribution of the retained mode shapes can be removed from the flexibility matrix forming the reduced flexibility matrix.

\begin{equation}
   F_i=K_i^{+}-\Phi_{i,r}\Sigma_{i,r}^{-1}\Phi_{i,r}^T
\end{equation}

The overall response is then approximated using a combination of the rigid body, free interface and attachment modes.

\begin{equation}
u_i\approx R_i\alpha_i+\Phi_{i,r}q_i-F_iB_i\mu
\end{equation}
The final dual CB reduction matrices can then be formed and the reduced global mass and stiffness matrices assembled.
\begin{equation}
    \begin{bmatrix}
    u_1\\u_2
    \end{bmatrix}
    \approx
    \begin{bmatrix}
    R_1&0&\Phi_{1,d}&0&-F_1B_1\\
    0& R_2 &0 & \Phi_{2,d}&-F_2B_2
    \end{bmatrix}
    \begin{bmatrix}
    \alpha_1\\\alpha_2\\q_1\\q_2\\\mu
    \end{bmatrix}=T
    \begin{bmatrix}
    \alpha_1\\\alpha_2\\q_1\\q_2\\\mu
    \end{bmatrix}
\end{equation}

\begin{equation}
    \tilde{M}=T^T\begin{bmatrix}
    \hat{M}&0\\0&0
    \end{bmatrix}
    T
    \quad   
    \tilde{K}=T^T\begin{bmatrix}
    \hat{K}&B\\B^T&0
    \end{bmatrix}T
\end{equation}

\begin{figure}[h!]
    \centering
    
    \begin{subfigure}[b]{0.48\textwidth}
        \includegraphics[height=58mm,trim={0 0 10mm 0},clip]{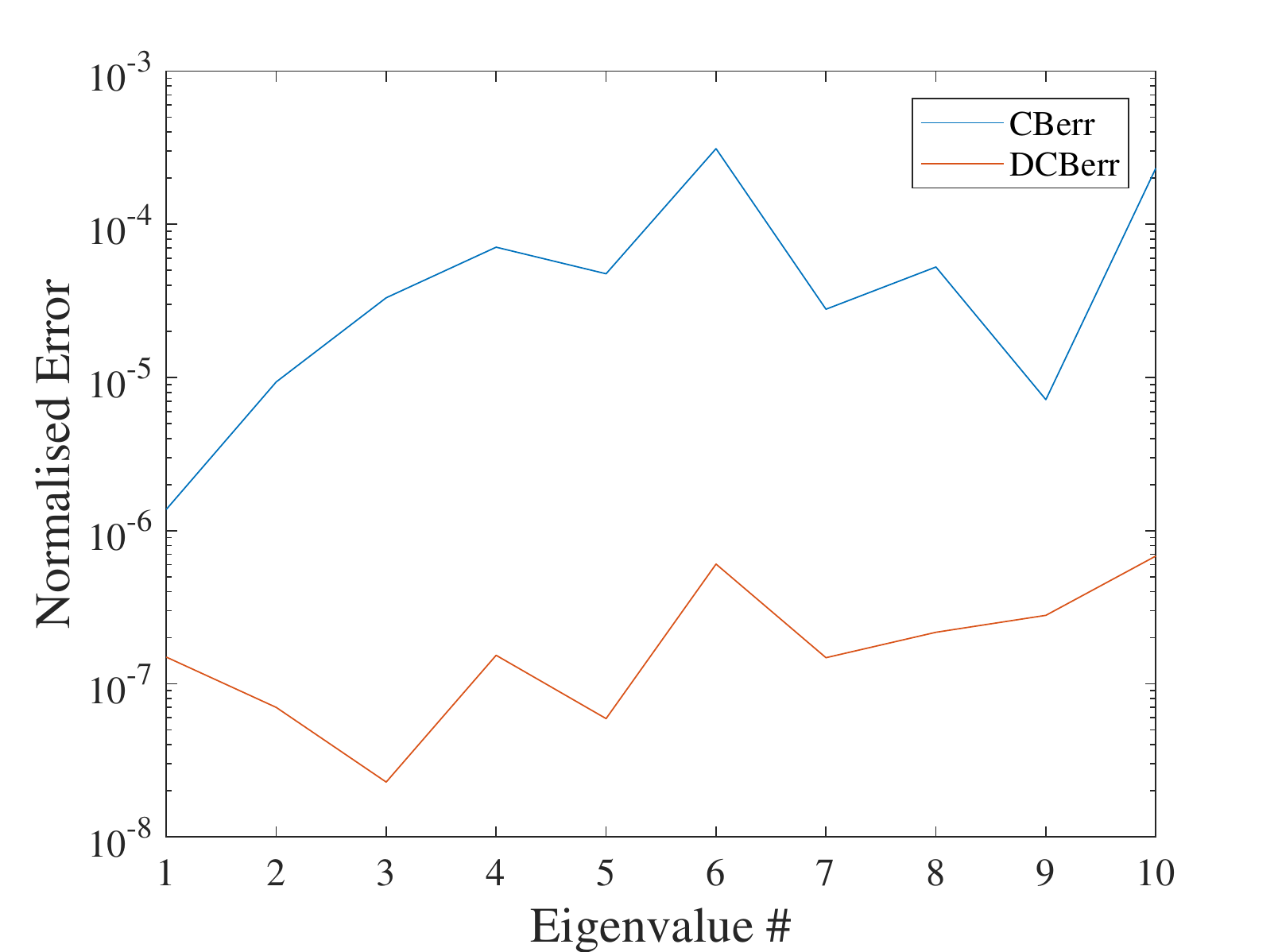}
        \caption{Error of first 10 eigenvalues}
    \end{subfigure}
    \quad
    \begin{subfigure}[b]{0.48\textwidth}
    \centering
        \includegraphics[height=58mm,trim={0 0 10mm 0},clip]{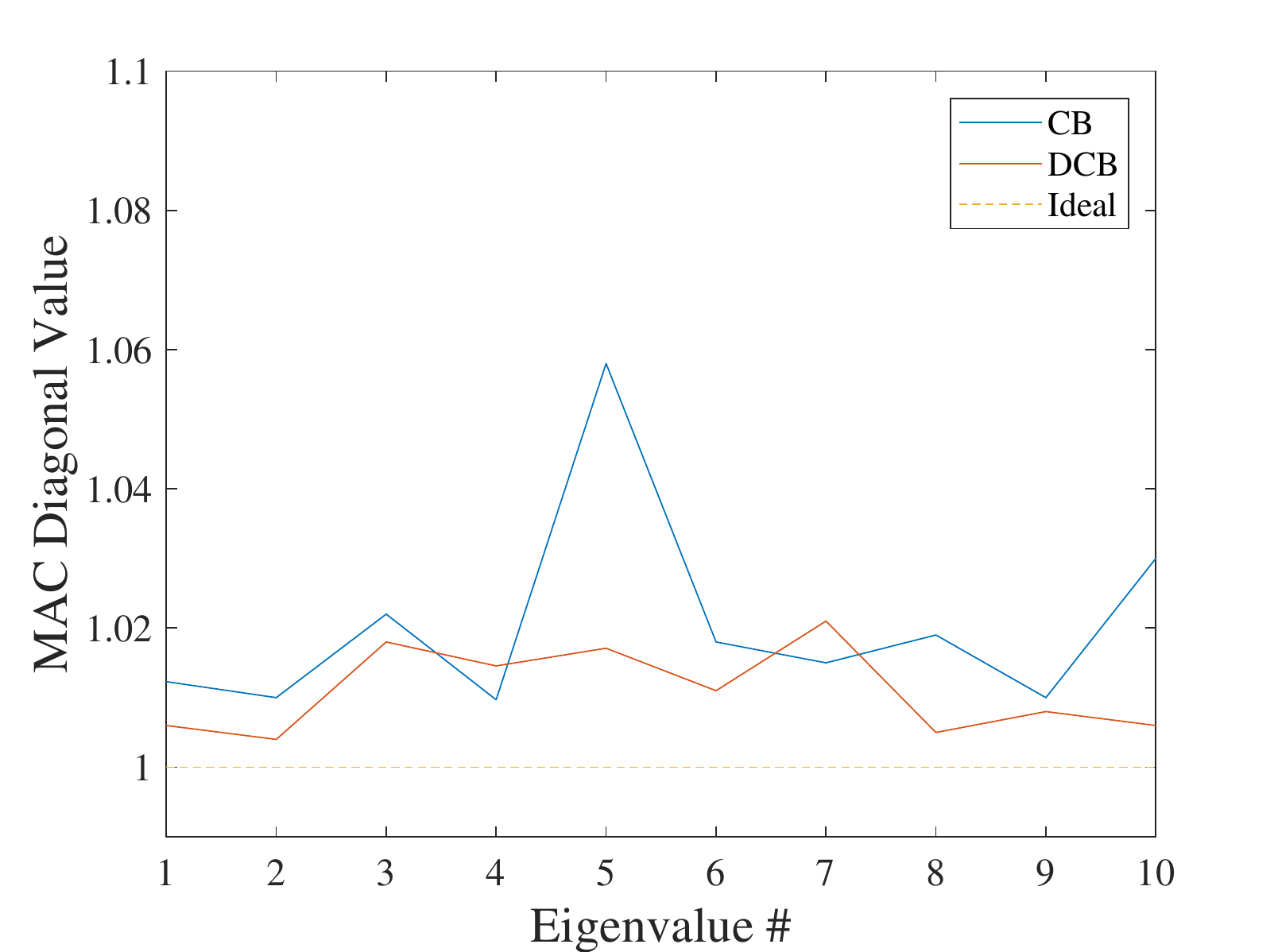}
        \caption{Diagonal value of MAC matrix}
    \end{subfigure}
 
    \caption{The error in the eigenvalues and MAC for the reduced systems compared to the full system.}
    \label{fig:eigs}
\end{figure}

Figure \ref{fig:eigs} demonstrates the performance of these two methods in preserving the eigenvalues and eigenvectors of the system. In both cases the same size of reduction matrix was used with, in each case, the number of DOFs reduced from the order of $3\times10^5$ to fewer than $500$. Further DOF reduction would be possible if interface reduction were also considered. It can be seen that for a given reduction size, the DCB method out performs the CB method in terms of eigenvalue fidelity. It is also evident that the DCB method seems to provide slightly better results with regards to MAC value as compared to the CB system.

%\begin{figure}[h!]
 %   \centering
  %  \includegraphics[width=75mm]{EigErr.png}
   % \caption{Error of the first 10 system eigenvalues from the CB and DCB reduced systems}
    %\label{fig:eigs}
%\end{figure}

\noindent \uppercase{\textbf{DS for Model Updating}} \vspace{12pt} 

In order to use these dynamic substruturing methods for model updating, their sensitivity to multiplication of the substructure stiffness matrix by a constant should be explored.
\begin{equation}
    K=\theta K_{0}
\end{equation}

With respect to the CB method, the fixed interface normal modes are first considered. By mulitplying a matrix by a constant value it can be shown that the eigenvectors of the matrix are unchanged, whilst the eigenvalues scale by the given parameter. As such, the fixed interace normal modes are unaffected by the parameteristion.

\begin{equation}
    M_{ii}^{-1}\theta K_{0,ii}=\Phi(\theta\Sigma)\Phi^T
\end{equation}
With regards to the constraint modes, it can be shown that the multiplication of the stiffness matrix by a constant also has no affect on the reduction basis.

\begin{equation}
    \Psi=(\theta K_{0,ii})^{-1}\theta K_{0,ib}=K_{0,ii}^{-1}K_{0,ib}
\end{equation}

For the CB method it is enough to update the substructural stiffness matrices whilst maintaining the same reduction matrices; this has previously been shown in \cite{Papadimitriou2013},\cite{Jensen2014}.

In the case of the DCB method, using the same reasoning as above leads to invariance of the free interface normal modes. The rigid body modes ,being modes not involving deformation, are also naturally unchanged by a variation in the stiffness matrix. However, with regards to the attachment modes, it can be seen that the parameterisation does bear an effect. 
\begin{equation}
   F=(\theta K_0)^{+}-\Phi_r(\theta \Sigma_{r})^{-1}\Phi_{r}^T=\frac{1}{\theta}F_0
\end{equation}

This results in an equivalent parameterisation of the reduction matrix although, this is very simply and with minimal computational effort execute in the reduction matrix.

\begin{equation}
    T_{\theta}=
    \begin{bmatrix}
    R_1&0&\Phi_{1,d}&0&-\frac{1}{\theta_1}F_1B_1\\
    0& R_2 &0 & \Phi_{2,d}&-\frac{1}{\theta_2}F_2B_2
    \end{bmatrix}
\end{equation}

\noindent \uppercase{\textbf{Results}} \vspace{12pt} 

In order to test the efficacy of both of the methods, 3 different test cases were considered. In two cases only one of the substructures had damage introduced whilst in the third damage was introduced simultaneously in both substructures.

\begin{figure}[h!]
    \centering
    \includegraphics[width=160mm]{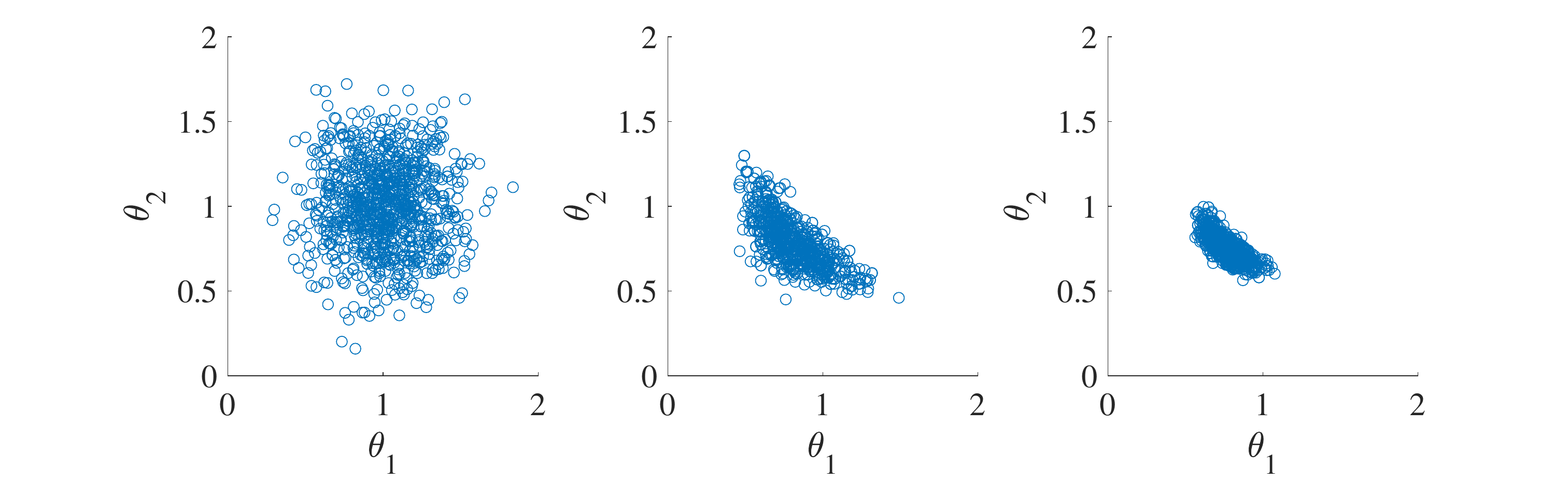}
    \caption{Scatter plots of the predicted variables in the 1st, 3rd and 5th stages of the TMCMC algorithm respectively.}
    \label{fig:tmcmc}
\end{figure}

Figure \ref{fig:tmcmc} illustrates how the parameter values coalesce towards the correct values over the course of the TMCMC algorithm. In this case the true parameters were set at 0.75 with the prior values set at 1. Over the course of the parameter identification the mean of the parameters approaches the true mean while, the variance of the prediction also decreases.

Table \ref{tbl:res} indicates that both the CB and DCB methods were able to accurately predict the damage parameters in all simulated experiments to a high accuracy. The overall performance of both algorithms was quite similar, and they were both able to correctly locate and quantify the damage.

\begin{table}[h!]
\centering
\caption{Results of the test cases for the CB and DCB algorithms}
\begin{tabular}{llllllll}
\hline
\multicolumn{2}{l}{Damage State} & \multicolumn{2}{l}{CB Estimate} & \multicolumn{2}{l}{DCB Estimate} & CB Error & DCB Error  \\

$\theta_1$   & $\theta_2$                        & $\theta_1$    & $\theta_2$                      & $\theta_1$    & $\theta_2$                       &          &            \\
\hline
1    & 0.75                      & 0.976 & 0.757                   & 0.988 & 0.759                    & 6.63e-04 & 2.97e-04   \\
0.75 & 1                         & 0.757 & 0.999                   & 0.756 & 1.020                    & 7.60e-05 & 4.50e-04   \\
0.75 & 0.75                      & 0.779 & 0.735                   & 0.779 & 0.743                    & 1.91e-03 & 1.53e-03  
\end{tabular}
\label{tbl:res}
\end{table}

\vspace{12pt}

\noindent \uppercase{\textbf{CONCLUDING REMARKS}} \vspace{12pt}

This comparative assessment demonstrated that both the CB and DCB methods can be used in the context of parameter updating for a linear FE model, and as a result, both locate and quantify damage. This comes with a huge reduction in computational time. Each of the parameter updating procedures completed 5000 sampling operations and required in the region of 1 hours run time. On the contrary, a single evaluation of the unreduced comprises an approximate duration of 10 minutes, meaning that an equivalent number of sampling operations would require several days of computing time.

Whilst both techniques performed similarly in the implemented test cases, it is worth considering that the final error in the eigenvalues of the identified systems was of the order of $1e-04$. In the case of the CB method, this is close to the error level achieved with perfect parameter values, as shown previously in Figure \ref{fig:eigs}. In the case of the DCB method however, this is around 3 orders of magnitude larger than the error found with perfect parameter values. A likely cause for this, may be attributed to the DCB updating procedure being halted by algorithmic convergence criteria, whilst higher accuracy may be possible with the DCB method given a greater number of samples.

\noindent \uppercase{\textbf{ACKNOWLEDGEMENTS}} \vspace{12pt}

\scalerel*{\includegraphics{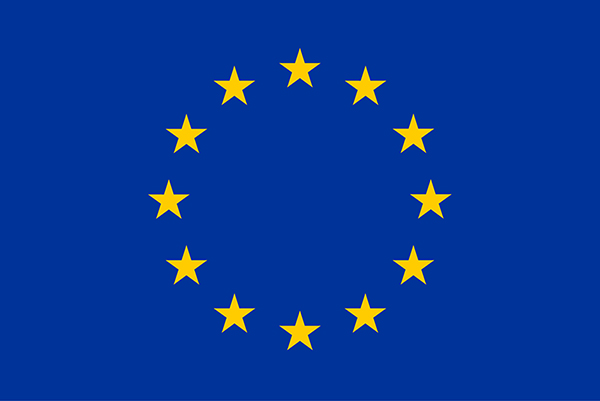}}{2*B} This project has received funding from the European Union’s Horizon 2020 research and innovation programme under the Marie Skłodowska-Curie grant agreement No 764547

%------------------------------------------------------------------------------------------------------------
%                                                 BIBLIOGRAPHY
%------------------------------------------------------------------------------------------------------------
\vspace{24pt}

\small 

\bibliographystyle{iwshm}
\bibliography{IWSHM_references}
\par\leavevmode
\end{document}